\documentclass[a4paper,12pt]{article}
\usepackage{graphicx}
\usepackage{infnprep}
\usepackage[top=2.5cm, bottom=3cm, left=3cm, right=3cm]{geometry}
\newcommand{\Header}{
  \resizebox{15cm}{!}{
  \begin{tabular}{rl}
  \includegraphics[width=5cm, trim={50 100 0 0}]{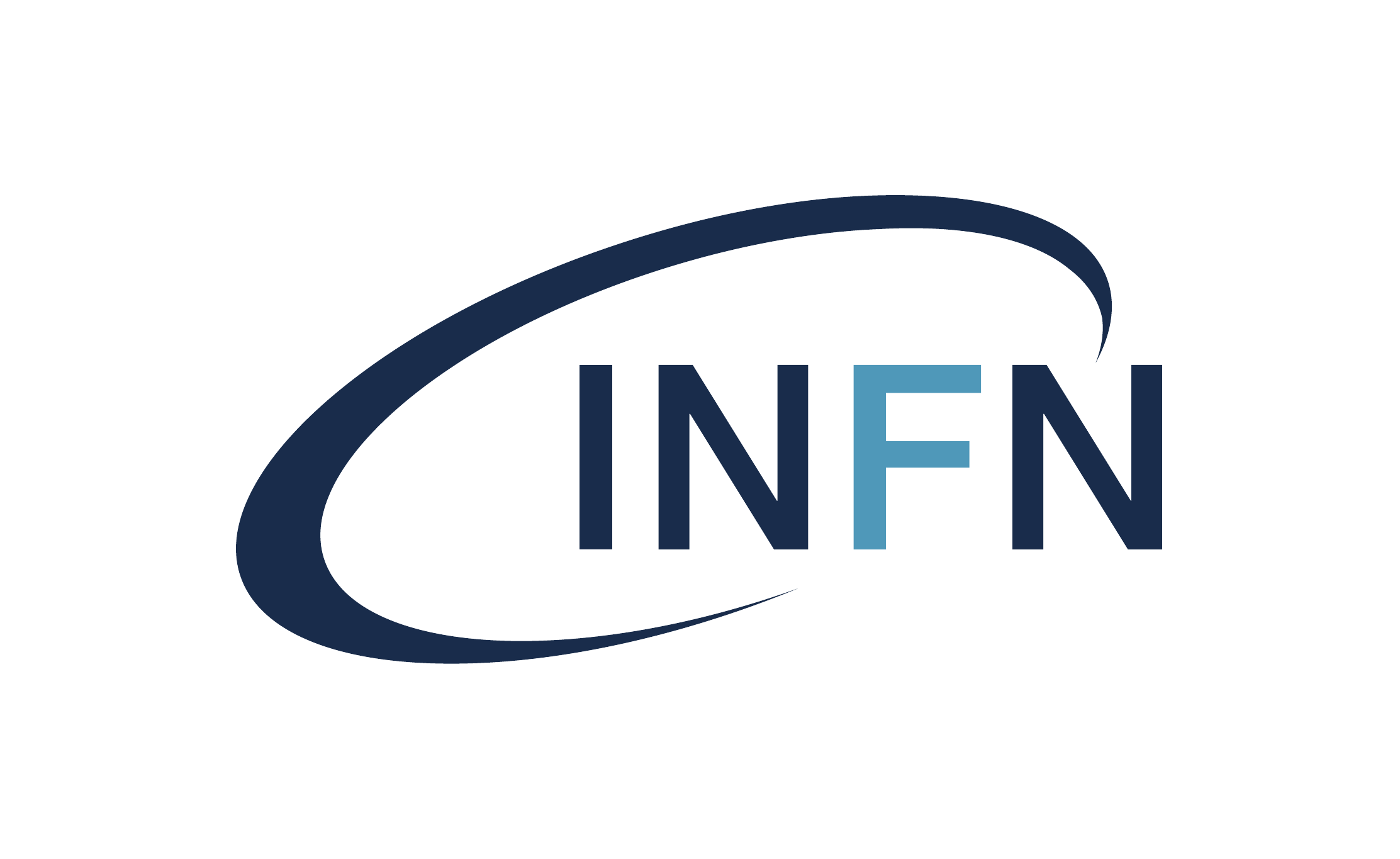} & {\LARGE\sffamily ISTITUTO NAZIONALE DI FISICA NUCLEARE}\\
      \\
  \end{tabular}
  }
\begin{center}
      {\large\sffamily Laboratori Nazionali di Frascati}\\
\end{center}
    \renewcommand{\arraystretch}{1}
\vskip 0.5cm
\rule{15.0cm}{0.09mm}
\vskip 1.5cm
  \begin{flushright}
      {\underline{\bf INFN - 18-05/LNF
       }}\\    % insert here the preprint number
      {\small\bf  25 maggio 2018} \\      % insert here the preprint Date
  \end{flushright}
  }

\usepackage[normalem]{ulem} %emphasize further-on italic 
\usepackage {ulem} %\emph{text}: text underlined, \sout{text}: text striked out 
\usepackage[english]{babel}
 \usepackage{hyperref}
\usepackage{natbib}
 \bibpunct[;]{(}{)}{,}{a}{}{;}
 \usepackage{type1cm}   
  
\hypersetup{ colorlinks=true,  linkcolor=blue,  filecolor=magenta,       urlcolor=blue, citecolor=black}

\usepackage{mathptmx}
\usepackage[intlimits]{amsmath}
\usepackage{helvet}
\usepackage{courier}
\usepackage{type1cm}         

\usepackage{makeidx}         % allows index generation
\usepackage{graphicx}        % standard LaTeX graphics tool
\usepackage{multicol}        % used for the two-column index
\usepackage[bottom]{footmisc}% places footnotes at page bottom
\def\BT{Bruno Touschek}
\def\RW{Rolf Wider\o e}
\def\PM{Pierre Marin}
\def\LAL{Laboratoire de l'Acc\'el\'erateur Lin\'eaire}
\def\FJ{Fr\'ed\'eric Joliot}

\def\FL{Fran\c cois Lacoste}
\def\JH{Jacques Ha\"issinski}
\def\ABL{Andr\'e Blanc-Lapierre}

\def\ENS{\'Ecole Normale Sup\'erieure}

\def\WW2{Second World War II}
\setcitestyle{notesep={, }}
\begin{document}
\begin{titlepage}

\title
  {\Header {\Large \bf Bruno Touschek and AdA: from   Frascati to  Orsay}
  %%%%liajune5%%%   
  \\ { \bf In memory of Bruno Touschek, who passed away 40 years ago, on May 25th, 1978}
  % The Franco-Italian collaboration and the first electron-positron collisions}
}

\author{
   Luisa Bonolis$^1$, Giulia Pancheri$^2$\\
{\it ${}^{1)}$Max Planck Institute for the History of Science, Boltzmannstra\ss e 22, 14195 Berlin, Germany}\\
{\it ${}^{2)}$INFN, Laboratori Nazionali di Frascati, P.O. Box 13,
I-00044 Frascati, Italy}
} 
\maketitle
\baselineskip=14pt
%%%\vskip 1cm %%%lia to have some space
\begin{abstract}
%This paper is dedicated to the memory of  Bruno Touschek, who passed away forty years ago, in Innsbruck, Austria, on May 25th, 1978.
The first electron-positron collisions in a laboratory were
 observed in 1963-1964 at the Laboratoire de l'Acc\'el\'erateur Lin\'eaire d'Orsay,
 in France, with the storage ring AdA, which had been constructed
 in the Italian National Laboratories of Frascati in 1960, under the
 guidance of Bruno Touschek. The making of the collaboration
 between the two laboratories included visits between Orsay and Frascati,
 letters between Rome and Paris, and culminated with AdA leaving Frascati
 on July 4th, 1962 to cross the Alps on a truck, with the doughnut
 degassed to $10^{-9}$mmHg through pumps powered by sets of 
 heavy
 batteries. This epoch-making trip and the exchanges which preceded it
 are described through unpublished documents and interviews with some of
 its protagonists, Carlo Bernardini, Fran\c cois Lacoste, Jacques Ha\"issinski, Maurice L\'evy. 
\end{abstract}

\vspace*{\stretch{2}}
\begin{flushleft}
% insert here the PACS number 
  \vskip 2cm
{ PACS: %\textcolor{red}
{01.65.+g, 01.60.+q, 29.20.-c, 29.20.db}}
\end{flushleft}
\begin{flushright}

  \vskip 3cm
 %%%lia \vskip 2cm
\small\it Published by \\
Laboratori Nazionali di Frascati
\end{flushright}
\end{titlepage}
\pagestyle{plain}
\setcounter{page}2
\baselineskip=17pt
\tableofcontents
\section{Introduction}
\label{intro}
In this paper we describe the making of the Franco-Italian collaboration which brought  AdA, the first electron-positron storage ring ever built, from  the Italian National Laboratories of Frascati to France, at the \LAL \ (LAL) in Orsay, where the linear accelerator could provide AdA with a source of  electrons  sufficiently intense to offer a good probability to observe collisions. AdA had been the brainchild of the Austrian born theoretical physicist Bruno Touschek, who came to Italy   in 1952, 
and proposed the construction of  a colliding beam machine to explore electron-positron annihilations, during an  epoch making seminar of March  1960. 

There are  accomplishments in science which, at first sight, appear to spring from  chance or improvisation, but  that, once inspected carefully, show the many different threads underlying  their development and the conscious decisions  taken all along, and which may  ultimately lead to   transform  pure research results   into society's gain.
 Such is the construction and development of particle colliders, which accelerate elementary particles such as electrons or protons, or their anti-particles, making them travel in opposite directions, within the same evacuated ring shaped chamber, called the doughnut because of its shape,  and then smash them one against the  other, and, through the observation  of the debris,   probe their internal dynamics or structure.  This type of accelerators  was envisioned  after World War  II, in Europe, in the United States, and in the USSR.  %\textcolor{red}
 {Their actual realization came through the contribution of different laboratories and scientists from many different countries --- Norway, Austria, Germany, Italy, the United States, France, Great Britain, all  these pathways eventually converging towards    the construction of the first particle-antiparticle collider in 1960  in Italy.} A comprehensive  description of  these various roads  is part of a larger work in progress we are preparing. What we shall do in this paper is to 
  to tell  the story of how two of these roads came to cross each other, leading to the demonstration of the feasibility of electron-positron colliders.
 
The first particle collider to    enter into operation used  electrons and their anti-particles, the positrons,  to smash one against the other, after a suitable number of them had accumulated. This first electron-positron collider  was  called AdA, an Italian acronym for  Anello di Accumulazione, storage ring in English. It was built in Italy in 1960, at the Frascati National Laboratories near Rome, and it was made to  achieve its potential to collide electrons against positrons three years later in France, at the  \LAL \ (LAL) in Orsay. Doubts had been raised as to whether   it could be possible to build an accelerator where   electrons and positrons could be observed to collide in the same ring. However,   when the word spread that AdA had started operating and, later on,  that collisions had been observed, projects for particle colliders were put in motion  in all the main particle laboratories in the world. 
 
AdA was a small machine, little more than 1 meter in diameter, which can still be seen installed   under a canopy, in one of the lawns    as one enters the Frascati Laboratories.  Since AdA's early  days, 
during the last  fifty years, scientists have built bigger and more powerful  colliders, the biggest of them all being  the CERN Large Hadron Collider (LHC). On the long road from AdA to the LHC, particle colliders have provided scientists with the tools  to control the dynamics of beams of elementary particles such as electrons, positrons, protons, anti-protons, heavy ions.  In the LHC, particles  travel  in a ring deep under ground, on  a 27 km long path, across France and Switzerland, passing along  the Cointrin airport of Geneva, in and out of the Jura mountains. With such a powerful machine, in 2012 CERN   could announce the discovery of the Higgs boson, the long sought missing link of the Standard Model of Elementary Particles. And, at the same time that colliders contributed to  a deep understanding of the structure of matter, it has been possible to   exploit this knowledge to built novel accelerators, which would be   powerful producers of  synchrotron radiation, currently used  in  a great variety of medical and technological applications.
 
The events which led to the construction of AdA in Frascati have been described in many publications, starting with Amaldi's biography \citep{Amaldi:1981} of \BT   --- the  prime mover and  the inspiration for   the construction of AdA --- to  the most recent
%\textcolor{red}
{testimonial of Touschek's life} in   \citep{Bernardini:2015wja}.
%%%lia%%%Bernardini:2016aa}. %\textcolor{red}
{Edoardo Amaldi, one of the founders of CERN, had invited \BT \ to come to work and do research in Rome in 1952. After Touschek's death in 1978, Amaldi prepared and published an extensive and unsurpassed description of Touschek's life and accomplishments. In Fig.~\ref{fig:TouschekAmaldi}, we show him  together  with Bruno Touschek in the 1960s.

 The transport of AdA to Orsay has also been described many times,  %\textcolor{red}
{and a preliminary outline} of the exchanges leading to the collaboration between Italian and French scientists was given in \citep{Bonolis:2011wa}. In this paper we shall present 
%\textcolor{red}
{a more complete reconstruction, with further details}, closing a number of gaps between the various letters and %\textcolor{red}
{visits and clarifying important events}. We shall make use of archival documents from both Sapienza University of Rome  and \LAL, as well of interviews with some of the scientists involved in  AdA's adventure. In particular, since 2003,  three  docu-films have been produced about \BT\ and AdA, namely  \href{https://www.youtube.com/watch?v=ami4kKkxSV8}{\it Bruno Touschek e l'arte della fisica}\footnote{\url{https://www.youtube.com/watch?v=ami4kKkxSV8}},   \href{http://www.lnf.infn.it/edu/materiale/video/AdA_in_Orsay.mp4}{\it Touschek with AdA in Orsay}\footnote{\url{http://www.lnf.infn.it/edu/materiale/video/AdA_in_Orsay.mp4}} 
and \href{https://webcast.in2p3.fr/video/60-ans-dexploration-de-la-matiere}{\it 60-ans-d'exploration-de-la-mati\`ere avec des acc\'el\'erateurs de particules}\footnote{\url{https://webcast.in2p3.fr/video/60-ans-dexploration-de-la-matiere}}, %\textcolor{red}
{which describes} 
%\sout{about}
 the impact of AdA on the development of electron-positron colliders in Orsay. Some of the quoted texts included in this paper have been extracted from these docu-films or from their unused parts.

 \begin{figure}
 %[htb]
  \centering
  \includegraphics[scale=0.6]{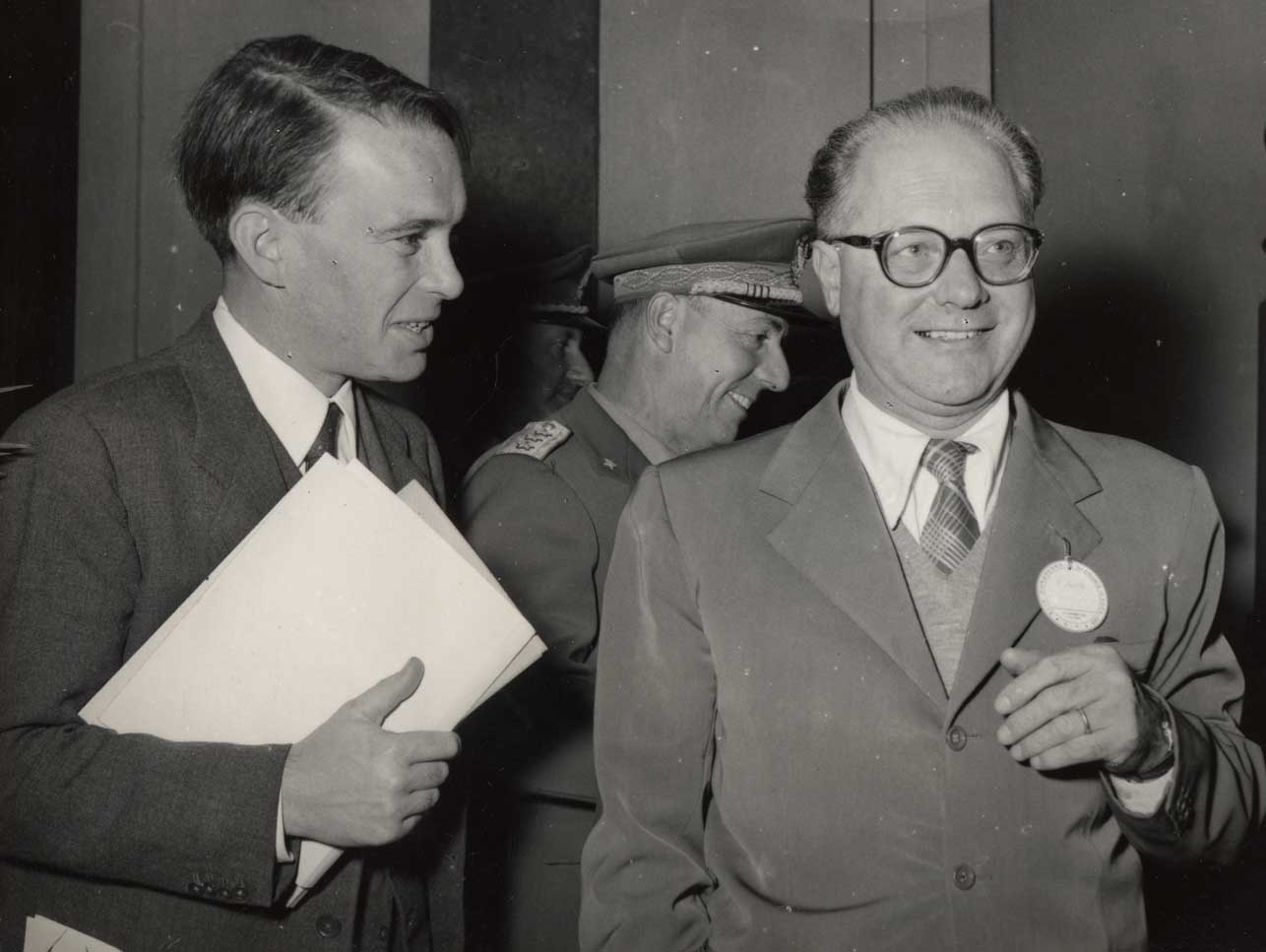} 
\caption{Bruno Touschek, at left, and Edoardo Amaldi at the beginning of the 1960s (Courtesy of Francis Touschek).}
 \label{fig:TouschekAmaldi}
  \end{figure}

The events of interest we shall describe  here  took place between June 1961 and July 1962, starting with a Conference in Geneva, where the 
Frascati team presented their electron-positron projects, up to   the time that AdA arrived in Orsay, around July 1962. 

As described in \citep{Bernardini:2004aa}, the AdA had been functioning in Frascati since February 27, 1961. 

%\textcolor{red}
The idea of 
%\textcolor{red}{building} 
electron-positron collisions had been suggested  by Bruno Touschek in a meeting held to discuss the laboratory future projects on February 17, 1960.\footnote{ %\textcolor{red}
{ An accelerator in which oppositely charged particles would smash against each other  in  head-on-collisions  had been first envisioned by \RW\ and patented on September 8th, 1943 \citep{Wideroe:1994}.} }
At that time, the electron synchrotron had been in operation at Frascati Laboratory since early 1959 \citep{Alberigi:1959} and, in November of the same year,  the CERN Proton Synchrotron (PS)  had accelerated its first protons. During the meeting, especially  devoted to the creation of a theoretical group in Frascati, Touschek suddenly came up with what, at the moment,  looked like a}  strange proposal: transforming the brand new electron synchrotron into  an electron-positron storage ring  \citep[449]{Amman:1989}.\footnote{See report of the meeting (L.N.F., Report N. 62, December 1960). See also B. Touschek, ``AdA e Adone'', manuscript, Edoardo Amaldi Archives, Sapienza University of Rome, Department of Physics, Bruno Touschek Papers (from now on Bruno Touschek Archive), Box 11, Folder  3.92.4, p. 7.}
This idea was rapidly discouraged. As Touschek later commented:\footnote{B. Touschek, ``A brief outline of the story of AdA'', excerpts from a talk delivered by
Touschek at the Accademia dei Lincei on May 24, 1974 (manuscript, Bruno Touschek Archive, Box 11, Folder 92.5, pp. 5-6).}
\begin{quotation}
%\textcolor{red}
{This proposal was not very tactful in front of a meeting of people who had built the machine and were proud of it and other who had spent years in preparing their experiments and were eager to bring them to a conclusion, to modify them and think of new ones to be carried out with the same machine.}
\end{quotation}
The project was saved by Giorgio Ghigo, Machine
Director of the Frascati Laboratories, who suggested to build a smaller ring, dedicated to  the storing of   electrons and positrons, and authored the blueprints of the magnet, inside which the AdA's doughnut would be placed. With this modification the project was enthusiastically approved by the Laboratory director, Giorgio Salvini.  Ghigo's suggestion was followed by  a detailed proposal presented by Bruno Touschek two weeks later, on March 7th, and approved shortly after with a 20 Million Lire budget. 

\begin{figure}[!]
\centering
\includegraphics[scale=0.69]{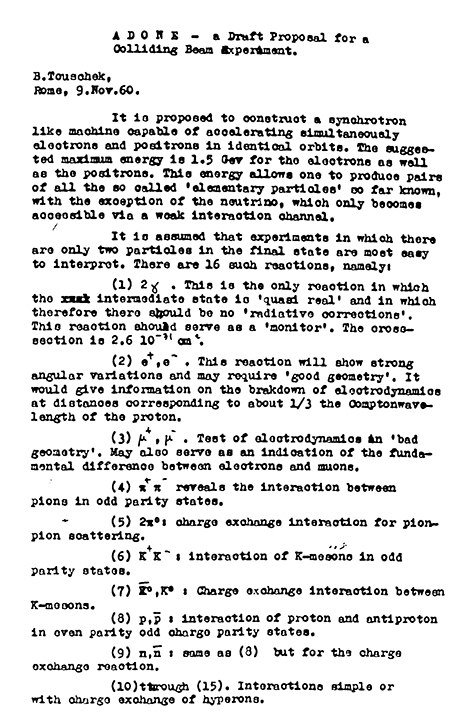}
%\figures/DSC00286-grafico-run-rosa}
\caption{Bruno Touschek's memo for Adone (Bruno Touschek Archive).}
\label{fig:AdoneProgetto}
\end{figure}

In November, it was clear that the ring would function and the first of four articles about the Frascati storage ring was submitted to the {\it Nuovo Cimento} and published  shortly thereafter \citep{Bernardini:1960osh}.  %\textcolor{red}
{That same November,} \BT \  wrote a memo  that envisioned the  building of a bigger ring, with a centre of mass (c.m.)  energy as high as $\sqrt{s}=3$ GeV (see Fig. \ref{fig:AdoneProgetto}).\footnote{ B. Touschek, `ADONE' a Draft proposal for a colliding beam experiment'' (typescript, Bruno Touschek Archive, Box 12, Folder 3.95.3, p. 3).} 

%\textcolor{red}

\begin{figure}[t]
  \centering
  \includegraphics[scale=0.8]{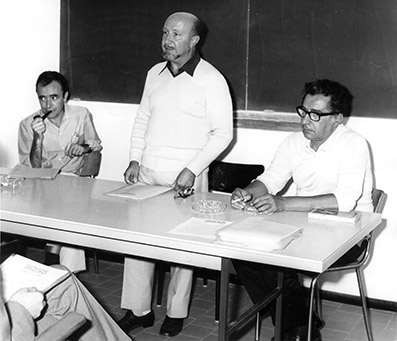} 
\caption{From the left, Fernando Amman,  Italo Federico Quercia and Ruggero Querzoli in the 1960s (Courtesy of LNF Archive).}
 \label{fig:AmmanQuerzoliQuercia}
  \end{figure}

{Following Touschek's memo on ADONE of November 1960, 
a document was prepared and submitted
% \sout{to INFN} \textcolor{red}
{for approval,} 
\footnote{
On January 27, 1961, F. Amman, C. Bernardini, R. Gatto, G. Ghigo and B. Touschek presented the Internal Report  
%\href{http://www.lnf.infn.it/sis/preprint/detail-new.php?id=3076}
``Anello di Accumulazione per elettroni e positroni (ADONE)'',
 LNF - 61 / 005, \url{http://www.lnf.infn.it/sis/preprint/detail-new.php?id=3076}.
} and in February 1961 a study group led by Ferdinando Amman was formally set up with the task of preparing a first estimate of the feasibility and costs of such a project.} By that time, a new director had come to the Frascati Laboratories, Italo Federico Quercia, shown in Fig.~\ref{fig:AmmanQuerzoliQuercia} with Amman,  head of the ADONE project, and Ruggero Querzoli, the 
senior experimentalist for the AdA group. 
%\textcolor{blue}{\bf Non so se esistesse questa gerarchia, ma sarei incline a dire di no. Chiediamo rapidamente a Carlo?Lia; Ghigo non era uno sperimentale, ne' lo era Carlo, dunque Querzoli era il senior experimentalist, direi.}

 ADONE, bigger than AdA, and  as beautiful as an Adonis, was planned to have such energy to produce the annihilation of the initial electron-positron pair into all the known particle-antiparticle final states,  such as pions or muons, i.e. $\pi^+\pi^-$, $\mu^+\mu^-$,  including the nucleons and their antiparticles, well above their production  threshold.\footnote{This maximum energy for the annihilating electrons and positrons  doomed Frascati not to be able to discover   the  charm-anticharm bound state, the $J/\Psi$, for which one needed   to reach more than of 3.1 GeV in the center of mass  \citep{Maiani:2017vde}. Once the discovery was announced \citep{Aubert:1974js,Augustin:1974xw}, the Frascati experimentalists forced ADONE  to reach the  higher energy value required to observe  the new particle, but could only confirm the discovery \citep{Bacci:1974za}, whose credit went to the two American groups, led by  Samuel Ting and Burton Richter, respectively. For a description of the days of the discovery, as seen from Frascati, see  M. Greco and G.P., "Frascati e la fisica teorica: da AdA a ADONE" ({\it Analysis}, issue 2/3 - June/~September 2008), \url{http://www.analysis-online.net/wp-content/uploads/2013/03/greco_pancheri.pdf} }
 %\textcolor{blue}{\bf LACKING REFERENCE} \citep{??}. }

\begin{figure}[htb]
\begin{center}
\includegraphics[scale=0.35]{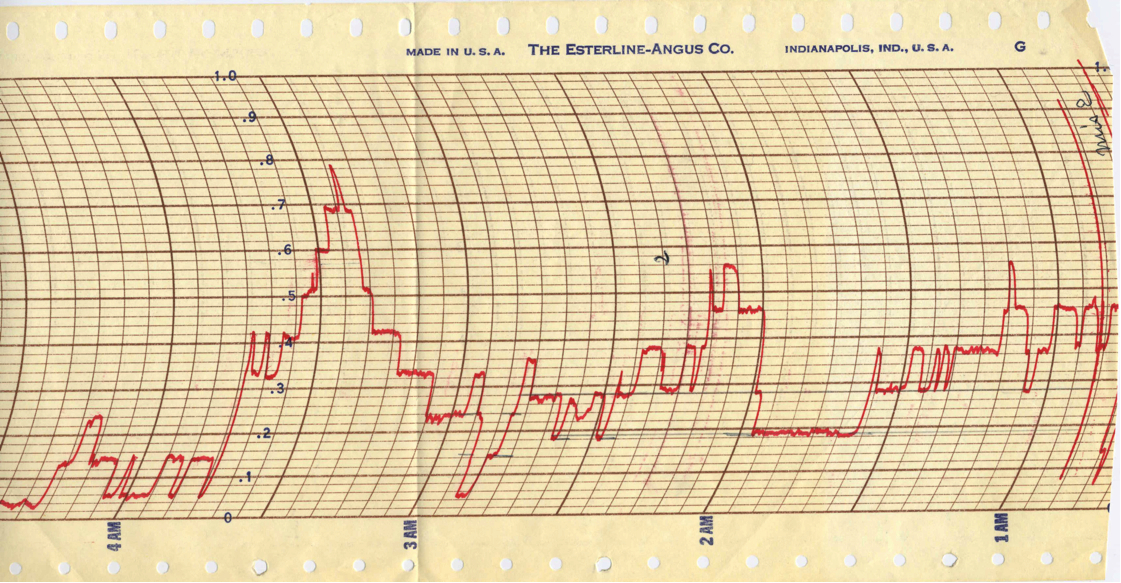}
\end{center}
%\figures/DSC00286-grafico-run-rosa}f
\caption{The phototube record of February 27, 1961, showing current versus time  steps that correspond to single
electrons entering or leaving AdA (courtesy of Carlo Bernardini).}
\label{fig:pinkoutputOscill}
\end{figure}

The enthusiasm was high in Frascati, when, in February 1961, the magnet was turned on and, less than a year after the official seminar where \BT \ had presented his proposal, the little ring started functioning: electrons and positrons  could be proven to circulate in the doughnut by observing   the light signal they emitted, the phenomenon also known as synchrotron light radiation. The  number of ``electrons''  circulating in the machine and the length of time these electrons would ``stay alive'',  was recorded.\footnote{It was never clear whether the light observed was emitted by an electron or a positron.} Part of the ammeter printout from the phototube record is  shown  in Fig.~\ref{fig:pinkoutputOscill}. 

%\textcolor{red}
{A similar recording  was sent by Touschek to Edoardo Amaldi on the morning following the second night of AdA's operation. During the next night Amaldi went to see with his own eyes something that nobody had ever seen before: the synchrotron light emitted by a \textit{single electron in orbit} which was visible to the naked eye through one of the portholes \citep[32]{Amaldi:1981}: 
\begin{quote}
Bruno took an immense pleasure in showing this phenomenon which, to a certain extent, was commonplace, but at first sight appeared incredible. His enthusiasm was extreme [\dots] 
\end{quote}

The  group  was now confident that AdA could lead  the way to higher energy physics, and  plans were put forward for a bigger and more powerful accelerator, one which would ``probe the quantum vacuum", and discover new particles, through  electron-positron annihilation. %\textcolor{red}

Spring 1961 saw more tests and measurements being done with AdA, while more theoretical work was completed by  \BT's colleague at University of Rome, Raoul Gatto, in collaboration with  Touschek's former student Nicola Cabibbo. They are both shown in Fig.~\ref{fig:GattoCabibbo}.  The first AdA paper had  aroused quite  some interest and Touschek was invited to  present  the Frascati work on storage rings  to the forthcoming International  Conference, to be held  in Geneva,  June 5th to June 9th,  1961.

\begin{figure}[htb]
\begin{center}
\includegraphics[scale=0.6]{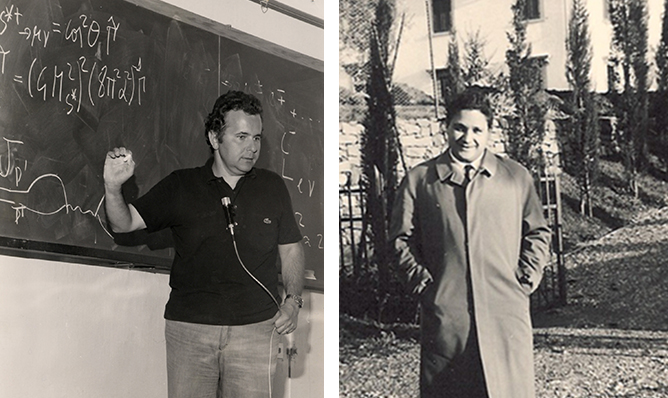}
\end{center}
%\figures/DSC00286-grafico-run-rosa}f
\caption{From the left, Nicola Cabibbo and Raoul Gatto in the early 1960s.}
\label{fig:GattoCabibbo}
\end{figure}

Throughout the following months, the AdA team  prepared   the contributions to
 the summer conferences, the now customary venue of presentation of the year's results, the first %\textcolor{red}
{being the one in Geneva}.   Our story starts there.

 \section{The 1961 Geneva Conference}

In Frascati, as summer 1961 approached, not everything was working as hoped. The main obstacle appeared to come from  the slow injection mechanism of particles from the electron synchrotron, too slow to produce a  number of electrons and positrons sufficient to prove that collisions between them were taking  place. Still the machine was operating and, in June, Touschek  presented these results to the CERN {\it International Conference on Theoretical Aspects of very high Energy Phenomena} \citep{Bell:1961gi}.\footnote{For these Proceedings, see also \url{https://cds.cern.ch/record/280184}.} His contribution appeared in the session on {\it Electromagnetic Interactions} which consisted of three talks, in the following order: one by Burton Richter,\footnote{In 1976, Burton Richter, from Stanford University, was awarded the  Nobel prize  for the discovery of a heavy elementary particle of a new kind, the $J/\Psi$, jointly  with Samuel Ting from MIT.} the second by Touschek, the third one by Raoul Gatto.\footnote{Raoul Gatto  was at the time at University of Rome, and, two years later, would   become Professor of Theoretical Physics at University of Florence.} Gatto's talk was focused on the theoretical aspects of electron- positron physics, presenting   two  papers authored with Nicola Cabibbo, a  short one published  in February 1960  \citep{Cabibbo:1960zza} and   a lengthier one submitted to  {\it The Physical Review} (recorded submission as of  June 8th,  1961),  just before leaving for the Conference \citep{Cabibbo:1961sz}. 
%\makebox{What observable means}

The contrast between the two first talks is striking. Richter seems to justify the choice NOT to start  electron-positron experiments, Touschek is announcing they have already built a small prototype and have proposed to build a bigger ring. Let us see the two talks in their most important passages, at a distance of almost 60 years, with what, in Italian, is called {\it Il senno di poi}.\footnote{ Wisdom, after the facts.} 

The session started with a talk by Burton Richter, from Stanford University, with the title {\it COLLIDING BEAMS EXPERIMENT }\cite[57]{Bell:1961gi}:
\begin{quote}
Let me begin by saying that we all hope that this will be the last talk about what we are going to do when the experiment is ready. We hope that next year we can talk about what we have done.
\end{quote}
And he then continues by describing the two ring set-up for the proposed electron-electron collision experiment and the hoped for time schedule. Then Richter goes on discussing {\it Positron Experiments}, and opens  with a denial:
\begin{quote}
Whenever this subject has been brought up in the past, we have refused to commit ourselves about its prospects. I am not going to change this policy, but I would like to discuss the difficulties of the positron experiments a bit.
\end{quote}
Having said this, he  %\textcolor{red}
{adds}:
\begin{quote}
These problems are not the main reason for our long-standing silence on the experiment. 
\end{quote}
Basically Richter insists that the electron experiments must first be shown to work well, before starting with positrons, namely
\begin{quote}
Until we know what we can do in storing a beam, we cannot say anything about the positron experiment.\end{quote}
Then Touschek starts,  with a simple concise sentence \citep[67]{Bell:1961gi}: 
\begin{quote}
Frascati is developing two storage rings.
\end{quote}
and then, after rapidly describing  the ADONE project, which was still under design, he moves to %\textcolor{red}
{describe} the first project:
\begin{quote}
The first project - AdA -  was started in February 1960. It was clear from the beginning that this project would be a gamble, the calculated intensity of the machine being about a factor 500 less than what was needed for experimentation. It was nevertheless decided to go on with the project, mainly because it was hoped that experience in storage problems could be most rapidly gained in this fashion and that eventually ideas for increasing the intensity might be forthcoming.  
\end{quote}
Touschek's contribution included many technical drawings, as well as   the phototube record of  the number  of electrons  circulating in AdA, shown in Fig.~\ref{fig:pinkoutputOscill}.
 In Fig.~\ref{fig:ymagnetAdA} we show two photos extracted respectively from Richter's and Touschek's talk.
\begin{figure}[!]
%[htb]
\centering
\includegraphics[scale=0.57]{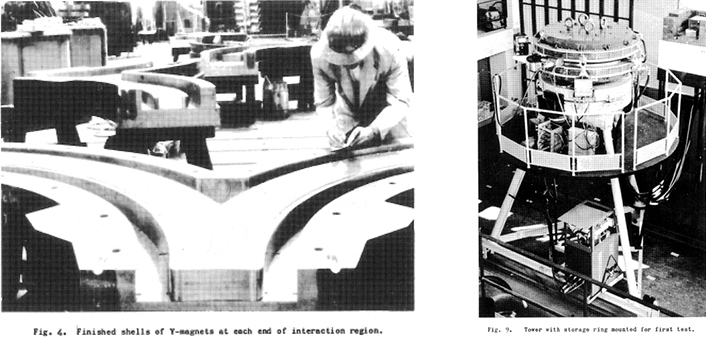}
%Y-magnet-Stanford-1961}
%\includegraphics[scale=0.3]{Ada-1961GenevaConf}
\caption{Two figures shown at the 1961 Conference, at left the Y magnet at Stanford, with the  region where the two electron orbits  are tangent to each other to produce electron electron collisions from  B. Richter's talk, at right AdA from Touschek's talk \citep{Bell:1961gi}.}
\label{fig:ymagnetAdA}
\end{figure}
In his talk, Touschek  gives one of the clearest expositions of AdA's technical details,   listing in his unique, and  extraordinarily precise way, all   the excellent reasons to start experimentation with electron-positron collisions.
 To complete this session, Gatto followed with a theoretical talk, where he  makes a very prophetic statement \citep[76]{Bell:1961gi}:
\begin{quote}
High Energy electron-positron colliding beam experiments may become a field of spectacular development in high energy physics.
\end{quote}

\section{And then came Pierre Marin and Georges Charpak: {\it Un vrai  bijou}}
\label{sec:marin}
The conference at CERN, and Touschek's talk, represent a landmark in the development of electron-positron colliders. Word spread that things were happening in the small Laboratory on the Tusculum Hills.

{In the far away Russian city of Novosibirsk,  beyond the Urals, at the Institute for Nuclear Physics, the scientists understood they were not alone in their work on electron-positron storage rings  and increased their efforts \citep{Baier:2008aa}. 
%How the world scientists  learnt of the Russian work will be the subject of chapter 7.  
Closer to Frascati, at CERN itself and in France, at the Laboratoire  de l'Acc\'el\'erateur Lin\'eaire d'Orsay, {\it le  LAL}, interest arose as to what the Italians were doing.  France, Italy and Switzerland are close,  travel between  the Laboratories was frequent,  and scientists could easily go and  see by themselves what was new. And this is precisely what happened:  two French scientists, one from Orsay, the other from CERN, went to Frascati to see with their own eyes\dots.

At the time,  in Orsay, a Linear Accelerator  had been  working since  1959, and  the question of  how to best exploit  its discovery potential  was often debated. The team which had built the Linear Accelerator included George Bishop and Pierre Marin, who knew each  other from Oxford  and had come together to France  in 1955, to work on the project. Marin was then 34 years old and    would  become one of the main artifices of France's accelerator program.  In 1961, with the linear accelerator by now successfully  built and working,  he was wondering  which  direction his research should take. When Marin asked around  for interesting things to do in his research, he was told to go see what was happening in Frascati. 

As Pierre Marin later wrote \citep[46]{Marin:2009}: 
\begin{quote}
Returning from a stay at CERN, after my thesis, I was searching for my own research directions, and G. R. Bishop suggested that I go to visit Frascati, where very intriguing things were happening. So, I went there, in the month of August, together with Georges Charpak, who was, at the time, collaborating with both CERN and American physicists on the measurement of the anomalous muon magnetic moment.\footnote{In 1992 Georges Charpak was awarded the Nobel Prize in Physics  ``for his invention and development of particle detectors, in particular the multi-wire proportional chamber''.}
\end{quote}
The visit actually took place in  July, as August is unbearably hot in and around the city of Rome.\footnote{Marin, in his recollections  almost forty years later,  indicates  the   month of  August for his visit. However  from \ABL's letter to Italo Federico Quercia in December 1961, we learn  that the visit had taken place in July.} 
Those who can,  escape to the  beaches, or go North, to the Dolomites,  or, in those days,  simply to the nearby  hills, the so called Alban hills, from the name  of the pre-Roman town  of Alba Longa. The Alban Hills, {\it I colli Albani}, include a half-circle of volcanic hills, which,  South East of Rome, limit  the plains where the city lies.  Among them, and closest to the city, is  the Tusculum Hill, sloping down to  the Frascati Laboratory mid-way to Rome,   cooler and  %\textcolor{red}
{shady} in the summer. In a hall next to the synchrotron Marin saw \citep[46]{Marin:2009} 
\begin{quote}
[\dots] a group of high caliber physicists [\dots] B. Touschek, C. Bernardini, G. Ghigo, F. Corazza, M. Puglisi, R. Querzoli and G. Di Giugno, [who] showed us with great pride a small machine, AdA, {\it un vrai bijou}\footnote{ A real gem.} [\dots]
\end{quote}
%\textcolor{red}
{In Fig.~\ref{fig:inostri}  %\textcolor{red}
{one can see} some of the Italian   protagonists of the AdA adventure. }

 \begin{figure}[htb]
  \centering
\includegraphics[scale=0.7]{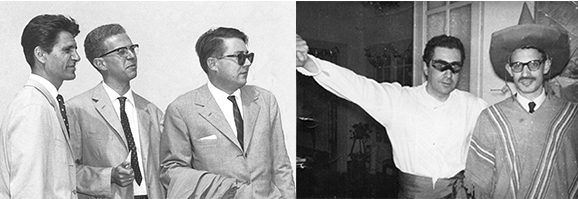}
%{QuerzoliDigiugno}\includegraphics[scale=0.73]{luce1}
\caption{
From the left, 
Gianfranco Corazza, Carlo Bernardini, Giorgio Ghigo, % \textcolor{red}
%In the right panel, we show 
and, in in  custom  party outfits,  Ruggero Querzoli and his student, Giuseppe Di Giugno (courtesy of Giuseppe Di Giugno.}%\textcolor{blue}{\bf Ma stavano insieme . il problema e' la scala della foto. le foto devono avere la stessa dimensione e poi essere messe alla stessa scala. Forse avevo un po' modificato anche quella di sinistra e te la rimando. Lia: se puoi fare mirror image dei 3 ma con meno spazio sopra la testa e pi' giacca di sotto, come vedi nella figura che ho fatto, possano evitare di cambiare la caption.}}.  
%AdA installed at Frascati Laboratories (\textcolor{red}{Credit?}).}
 \label{fig:inostri}
  \end{figure}

But,  as Touschek and Bernardini knew,  there were problems with their wonderful little machine. The group of Italian and French scientists started thinking about  ways to make AdA produce some  physics, beyond the  great pride of having succeeded in accumulating electrons and positrons in some  number. The estimates which Touschek had done in March  of the previous year, and which were  based on  cross-section calculations  by his colleagues Nicola Cabibbo and Raoul Gatto, showed that, even if some electrons and positrons were  circulating in the ring, luminosity was too low and annihilation could not be proven. To go beyond and demonstrate the feasibility of this type of machines to do physics at high energy, the problem of injection had  to first  be  solved.  The Italian team had applied some very ingenious ideas but the rate  remained  orders of magnitudes smaller than what a successful experiment would require. The problem was always the same: the beam of electrons extracted from the synchrotron was too weak, it could not lead to a photon flux high enough  to create enough  positron electron pairs in the AdA inner target.    Then,  Pierre Marin observed   that, in Orsay, the Linear Accelerator provided a well focussed  500 MeV electron beam,  with an excellent  intensity of 1 microamp\`ere. Bernardini and Touschek looked at each other,  moved slightly  aside, talked for a few minutes, and then came back to Marin. The question they posed was \citep[47]{Marin:2009}:
\begin{quote}
`` [\dots] Do you think that LAL would agree to receive AdA?"\\
I replied: ``A priori, the  new director \ABL\ would be quite open to welcome this sort of ideas.'' 
\end{quote}
%\textcolor{red}
{In Fig.~\ref{fig:BT-Marin}, we show Bruno Touschek and Pierre Marin in the 1960's.}

Hopes were revived and unexpected perspectives opened. Marin and Charpak left. The seed of the future of AdA as a feasible way to high energy colliders had been planted.  Marin carried back to Orsay  a promise to follow up the idea of a collaboration and perhaps a transfer of people and machinery to France. In the months to follow, steps for a  collaboration and the transfer of AdA from Frascati to Orsay were  put in motion. 

  \begin{figure}[htb]
  \centering
  \includegraphics[scale=0.8]{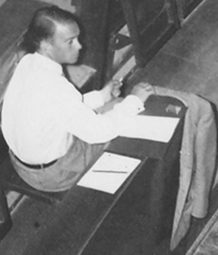}   \includegraphics[scale=0.8]{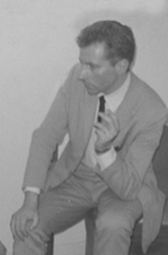}
\caption{Bruno Touschek, at left, and Pierre Marin, at right,  in the early 1960s. \BT's photo is courtesy of Giuseppe Di Giugno, Marin's photo is courtesy of \JH. }
 \label{fig:BT-Marin}
  \end{figure}

The operation was not trivial, as it involved moving  a working  accelerator,  which was property of a Government institution, 
across two countries, a trip of about  1500 kilometres, through custom and border controls. We should remember that in 1961, in Europe, there was no free circulation of goods and people  and AdA was precious: it had costed some 20 million Lire,  at the time a large expenditure for pure research, in a country  as poor as post-war Italy. The exchange therefore needed to be approved at the highest level. In addition, as the exchange of letters between the two Laboratories later  showed, the collaboration involved a matter of scientific policy in Europe, something to which Edoardo Amaldi, one of CERN's founders, was particularly sensitive.

\section{The 1961 Conference at Aix-en-Provence}
%paragraph{THE 1961 CONFERENCE at AIX-EN-PROVENCE}
%he 1961 Conference at Aix-en-Provence}
As soon as the August vacations ended, both in Orsay and Frascati, activities started in earnest:  the immediate scope was preparation of talks to present at  the imminent conference in Aix-en-Provence, where  the community involved in nuclear and particle physics, both experimentalists and theorists, was to gather  in mid-September from 14th to 20th . The title of the conference, {\it The Aix-en-Provence International Conference of Elementary particles},  addressed directly, for the first time, the emergence  of the field of elementary particle physics, singling it  out from nuclear  or accelerator or high energy physics, where  it had been so far included. It was mostly a conference where up-to-date theoretical ideas would be  debated in the plenary sessions, with experimental talks and other theory papers presented in the parallel sessions. The Italians were presenting  results from experimentation with the Frascati electron synchrotron, Ruggero Querzoli being the team leader and  giving the talk. From Orsay,  Marin was attending,  having received instructions from LAL director, Andr\'e Blanc-Lapierre, to probe the Frascati colleagues about both AdA and ADONE. 

 The concluding remarks were given by  Richard Feynman. It is worth repeating some of them here, applicable as well to what was happening at the Conference \citep{Freynman:1962nha}:
\begin{quote}
I want to ask what is most characteristic of the meeting --- what new positions are we in at the present time --- what kind of things do we expect in the future?

At each meeting it always seems to me that very little progress is made. Nevertheless, if you look over any reasonable length of time, a few years say, you find a fantastic progress and it is hard to understand how that can happen at the same time that nothing is happening (Zeno's paradox).

I think it is something like the way clouds change in the sky --- They gradually fade out here and build up there and if you look later it is different. What happens in a meeting is that certain things which 
were brought up in the last meeting as suggestions come into focus as realities.  They 
drag along with them  other things about which a great deal is discussed and which will become realities in focus at the next meeting. 
\end{quote}

 %\textcolor{magenta}
 {Thus, nothing was happening, apparently, as Feynman says, but he was (of course) right,  as new realities would soon come into focus and ``fantastic progress" would take  place in due time.} 
 
 %\textcolor{magenta}
 {Among the 90 talks, both plenary and parallel \citep{Cremieu-Alcan:1961qea}, there was one by  Raoul Gatto, 
 %\citep{Cabibbo:1961sz}, 
  who had co-signed ADONE's proposal in the previous month of February.
   %\sout{{\bf check}.} 
  His talk was in one of the parallel sessions  and, as he had done  in Geneva two months before, it   addressed the discovery possibilities of electron-positron physics.  } 

 %\textcolor{magenta}
 {In addition to Gatto's talk, detailed discussions, about ADONE's prospects and AdA's results, took place between Pierre Marin  and  Ruggero Querzoli from Frascati. Querzoli   at the time was   working on the synchrotron, but also  on  AdA, and  had been one of the {\it physiciens du cru}, whom Marin had met in July,  when visiting AdA. }
 
 %\textcolor{magenta}
 {Feynman could not have known of such discussions, and there is no mention of 
 % no notice was given to 
 Gatto's talk in his  summary. This is partly understandable in light of the fact that the calculations by Cabibbo and Gatto were not using new techniques or envisioning new theoretical scenarios: they actually used the tools of QED and  Feynman graphs to calculate the cross-sections of all known processes of interest in  electron-positron physics, some of which were already present in the literature. The value of the paper was in the exhaustive study and  its completeness. The great novelty of the paper, which had  been submitted the previous  July to {\it The Physical Review} \citep{Cabibbo:1961sz}, was that it presented realistic calculations for  processes that could now be observed and  measured in presently foreseeable experimental set-ups, such as ADONE's. }

%\textcolor{magenta}
{And so, while ``nothing was happening'',  our \textit{heroes} were building in the wings and  a revolution was being set in motion which would  produce new powerful tools of discovery.}

\section{Letters and visits}
%were exchanged}

%\paragraph{LETTERS WERE EXCHANGED}
On his return  to Orsay, Marin prepared a detailed report to his director,  accompanying it with a hand written note. The report, {\it Entrevue avec le Professor Querzoli de Frascati  le 19-9-61} is divided into two parts, the first concerns AdA, the second is about ADONE.\footnote{Copies of the report and of the accompanying letter by Pierre Marin were obtained from LAL Archives, courtesy of Jacques Ha\"issinski.}
About AdA, after describing the present state of the machine, successes, and  limitations,  such  as only a moderate  vacuum, a slow injection mechanisms etc., Marin listed the expectations of the Frascati group in case that  a much better vacuum, i.e. $10^{-9}$ mmHg  {\it vs.} the actual $10^{-6}$, could be reached. Among these, were the measurement  of the annihilation process $e^+e^- \rightarrow 2 \gamma$, studying the phenomenon of space charge and   the effects of changing the beam energy. But then, at the end of this list of expectations, comes the crucial proposal. We transcribe it here in its original French version:
\begin{quote}
Si les pr\'evisions de calculs sont exactes, il semble possible de r\'ealiser ce programme \`a Frascati [\dots] S'il s'av\'errait qu'il ne puisse \^etre r\'ealis\'e \`a Frascati, A.D.A. serait transport\'e \`a Orsay aupr\`es de l'Acc\'el\'erateur Lin\'eaire.\footnote{If the predictions of the calculation are exact, it seems possible to realise this program in Frascati \dots If it will happen that it cannot be realized   in Frascati, A.D.A  will  be transported to Orsay, next to the Linear Accelerator.}
\end{quote} 

The report then addresses the new Frascati project, ADONE, an electron-positron storage ring, with a beam energy of 1.5 GeV.  Marin informs his director that, although the project has not been officially approved, the Italians are rather confident it will be accepted, notwithstanding its much higher cost with respect to AdA. ADONE's budget was 2.5 billion lire, two orders of magnitude costlier than AdA's, with half of it  to be spent on a powerful linear accelerator. This was a very ambitious project, which could be started after some further experimentation with AdA. Marin adds  that Querzoli  is very favourably inclined to have   a French scientist  
visiting with the AdA team,  in the coming months.

\paragraph{A STORMY BEGINNING}
%\subsection{A stormy beginning}
Three months passed however, and no further exchanges seem to have taken place in the immediate period following this report. Then,  at  the end of December,  a letter from \ABL \ reached Italo Federico Quercia, then director of the Frascati Laboratories.\footnote{See also  Quercia's biography, {\it Italo Federico Quercia, note biografiche  e documenti},  edited by  Ugo Spezia, Collana di Storia della Scienza, Patrocini : SIF, ENEA, INFN, 2007.} The letter, dated December 22nd, 1961, starts with a rather unexpected sentence: ``Dear Professor Quercia, we are starting preliminary studies for  a 1.3 GeV storage ring for positrons and electrons in Orsay", and then  continues proposing that two or three people from Orsay come to Frascati to discuss some points about storage rings. \ABL \ proposed that visits  could  start any  time after  the 23rd of January.\footnote{LAL Archives, courtesy of Jacques Ha\"issinski.}

This letter is very interesting, as it indicates  that   a  French decision, to start planning for  a storage ring at LAL, must have  been taken between the time of Marin's report to  the Orsay director on September 1962,  and the December letter to Frascati,  proposing  a visit by some French scientists.  Edoardo Amaldi answered with a letter which reached Orsay one month later, and which was written in French, which, in those days,  was the official language of diplomacy.\footnote{LAL Archives, courtesy of Jacques Ha\"issinski.} Indeed,  \ABL's  letter had ruffled some feathers in Rome, as we shall see next.

\ABL's letter was probably received in Frascati during the end of the year vacation. As soon as the normal laboratory activities started again, Quercia called a meeting with Fernando Amman, Carlo Bernardini, Gianfranco Corazza and Bruno Touschek for Wednesday January 3rd. He   showed them the letter from the French director  and   the decision was taken to answer positively to the French request, including agreeing to the  proposed date of January 23rd  for the visit to start.\footnote{I. F. Quercia  to B. Touschek, January 12th, 1962, Bruno Touschek Archive, Series 1, Folder. 4, Box 1, 1962-67 Correspondence.}
% (now La Sapienza \textcolor{blue}{\bf ma non si chiamava prima La Sapienza? Adesso e' Sapienza Universita' di Roma \url{https://www.uniroma1.it/it/pagina-strutturale/home}}) 
%{\textcolor{blue}{Perche' non c'erano Querzoli e/o  Ghigo?}}  
Hopes for a collaboration which could solve AdA's injection problem and open the way to prove full feasibility of electron-positron storage rings, were renewed. Touschek, among the AdA team, was particularly anxious that a positive answer be sent immediately to  the French colleagues. A letter by Quercia to Touschek, dated January 12th, less than 10 days after the meeting, implies some tension between Quercia and Touschek on this issue, since Quercia mentions a telephone call by  Touschek "this morning",  whose reason  he "could not understand well".\footnote{I. F. Quercia to B. Touschek, January 12th, 1962, Bruno Touschek Archive,  Series 1, Folder. 4, Box 1, 1962-67 Correspondence.} We can only imagine that Touschek,  when he saw that no letter to LAL had been sent after almost ten days from the meeting and more than two weeks from \ABL's letter had passed,  became upset, fearing that the French would be moving ahead with their proposed plan (for ACO, Anneau de Collisions d'Orsay), and that the visit would be delayed. {He must have pressed  the Laboratory director   for an answer, in not too diplomatic terms. This is quite understandable, since} he certainly knew that the future of AdA rested on the collaboration with  LAL scientists and the use of their Linear Accelerator.
 
 Touschek's phone call to Quercia did produce a reaction, and, on January 16th, Amaldi, then director of the Physics Institute of University of Rome, and INFN president, answered to \ABL.\footnote{For a list of INFN Presidents, see \url{https://www.presid.infn.it/index.php/it/10-articoli-del-sito/73-presidenti-dal-1954}.} The matter, indeed,  did not simply involve  a friendly agreement among scientists and two laboratories,  as Touschek may have thought. Its complexity can be glimpsed from  the fact that Quercia had sent a formal letter to  answer  Touschek's phone call. Since the University of Rome and the Frascati Laboratory were connected by a direct  shuttle bus, which used to leave from the University of Rome every hour, 
 % \textcolor{blue}{\bf to check if every hour or twice a day, a me sembra  che  fosse ogni ora},  
 it would have been  simpler for Touschek and Quercia   to just   talk to each other:  instead we have a formal letter, from the Laboratory director, in Frascati,  to Touschek, in Rome. In fact, as one can see  from the correspondence which  followed,  the matter   needed to be thought over,  with adequate consideration of scientific priority.
 
  The reason for  the delay in answering \ABL's letter  is apparent in the  response  which Edoardo Amaldi sent to \ABL , four days later, on January 16. Amaldi starts his letter with the acknowledgment of the proposed  collaboration, referring to Amman and Touschek for its implementation,  but then   he addresses   the point of major interest for him, that of  European accelerator strategy and Italian scientific priority. Unlike the previous one by \ABL \  or   anyone of successive exchanges, this is a formal letter, as proven by  the fact that it was     written in French, the language of diplomatic exchanges, not in English.   We quote here   from the letter:\footnote{LAL Archives, courtesy of Jacques Ha\"issinski.} 

 \begin{quote}
 [\dots] Pour ce qui concerne la construction d'un Acc\'el\'erateur qui accumule dans le m\^eme anneau \'electrons et positrons, je d\'esire vous faire savoir que le group dirig\'e par l'Ing. Amman est d\'ej\`a arriv\'e \`a un degr\'e plut\^ot avanc\'e dans le projet d'une machine de ce genre pour 1,5 + 1,5 GeV [\dots] et que cet appareil constituera la partie essentielle de notre Second Plan Quinquennal [\dots].
 
 Nous croyons, pourtant, qu'il faudrait, en pr\'eparant  des autres programmes, consid\'erer tout cela, afin d'\'eviter des redoublements [\dots]
 Ceci est un cas particulier du probl\`eme plus g\'en\'eral de la coordination [\dots] des plans de recherche de divers groupes nationaux surtout entre deux pays d\'ej\`a si li\'es par des int\'er\^et[s] communs, comme la France et l'Italie.\footnote{%\textcolor{red}
 {Concerning the construction of an Accelerator  that can store  in the same ring electrons and positrons, I wish to let you know that the group directed by Eng. Amman has  already reached an advanced stage in planning  such machine with [energy] 1,5 + 1,5 GeV [\dots] and this apparatus will constitute the main part of our Second Five-Year Plan [\dots]. 
  We believe, therefore, that, in starting other  programs, one should consider all this, in order to avoid duplications [\dots] This is a particular case of the more general problem of coordination [\dots] of the research plans of different national groups, especially between two countries already so closely connected by common interests such as France and Italy.} }
     \end{quote}
Amaldi then continues with the authority given to him as one of the founders of CERN and protagonist of  the European scientific reconstruction after the war: 
     \begin{quote}
     Un premier pas, dans ce sens, a  \'et\'e d\'ej\`a fait avec la construction de l'European Accelerator Study Group, aux r\'eunions duquel  on a present\'e le project italien d\`es Decembre 1960.\footnote{
     A first step in this direction has already been taken with the establishment of the European Accelerator Study Group,  where the Italian project has been presented since December 1960.}
          \end{quote}
    { This letter was received in Orsay, on January 24, and was followed by another letter  by Amaldi to \ABL, in which, having clearly  stated  Italy's priority, Amaldi  welcomed the start of the collaboration. In this second letter, dated January 23rd,  the exchanges between the two laboratories were blessed and specific plans for visits could start. Amaldi invited \ABL \  to come to Frascati  ``accompanied, eventually, by another Member of your Group". The occasion was going to  be an informal meeting held on the dates 7-8-9 of February, to discuss present results from the electron synchrotron, work at CERN by Italian groups, and, last but certainly not the least in Amaldi's priorities, reports on the ADONE project.  This second letter reached Orsay on January 29th.\footnote{LAL Archives, courtesy of Jacques Ha\"issinski.}
 At  his end, after his December letter, \ABL\ had not remained idle: even before receiving  the Italians' answer to his query, he had contacted the  French Atomic Energy Commission, {\it Commissariat \'a l'\'Energie Atomique (CEA)}, probably in the person of its President, Francis Perrin.\footnote{Amaldi's letter  reached Orsay only on January 24th, but the French Atomic Energy Commission,  is mentioned in  a  16th January letter written from Orsay to Quercia ( on the same day as Amaldi's positive answer for a collaboration left Rome). We have not located this letter, but its existence is acknowledged, and its content described, in a January 23rd letter by Quercia to Andr\'e Blanck-Lapierre --- director to director.} We learn that, on January 16th, \ABL \ had proposed  the new date of February 5th 
    %( rather than the original January 23rd)
      for a visit by three French scientists, F. Fer, Marin and Boris Milman, adding that  one or two people from the French Atomic Energy Commission could  be part of the expedition.}\footnote{Marin and Milman had been part of the team which came from Oxford in 1955. F. Fer is among the participants to the Geneva 1961 Conference.} The inclusion of CEA %official 
 visitors
 %, not just scientists,
     indicates the importance the French scientific establishment attributed to the storage ring projects. 
%\subsection{Visits and encounters: towards the transfer of AdA}
 \paragraph{VISITS AND ENCOUNTERS: TOWARDS THE TRANSFER OF ADA}
  From February through  June 1962,  the collaboration was put in motion, resulting in  reciprocal visits % \textcolor{red}
  {and (at least) one  meeting in Geneva. }
  
  The first visit took place, as proposed  by Amaldi, on the occasion of the Frascati {\it Congressino}  held from  7th to 9th  February 1962. Amaldi's introductory talk was dedicated for the major part to the ADONE project, which  he clearly saw as crucial for the future of Italian and European high energy physics. \ABL  \ did not come, but four French scientists were welcomed by Amaldi in his opening  address  to the participants,  H. Bruck, F. Fer, F. Guerin (M.le Guerin, in fact, offers a welcome appearance of a woman in this story) and P.   Marin.\footnote{The Report of the Meeting is available at \url{http://www.lnf.infn.it/sis/preprint/detail-new.php?id=2980}, as LNF-62 /  037.}
 % , but reference needs to be prepared. }
  In addition, the non Italian participants included  F. Lefran\c cois, also from Orsay and Richard Wilson from Cornell. We notice  a   UK scientist from Harwell, P.G. Murphy from Rutherford Laboratory, attending the meeting. This may have created an opportunity  for Touschek to visit the UK, later in February.\footnote{This visit  is mentioned in  a letter from Touschek to Ernest Rae, at Harwell. For this and  
  %to be found as  TOUSCHEK\_ARCHIVIO\_CORRISPONDENZA. \textcolor{blue}{\bf perche'  le maiuscole qui?}
  other earlier contacts,  such as a letter from O'Neill dated February 8th, 1962, see  Edoardo Amaldi Archive, Sapienza University of Rome, Bruno Touschek Archive,  Series 1, Folder. 4, Box 1, 1962-67 Correspondence.} A thank you note from F. Fer to Touschek is shown in Fig.~\ref{fig:FerToTouschek-Rae}, together with a letter from Touschek to E.R. Rae from Harwell.
  
   \begin{figure}[t]
   %[htb]
  \centering
  \includegraphics[scale=0.17]{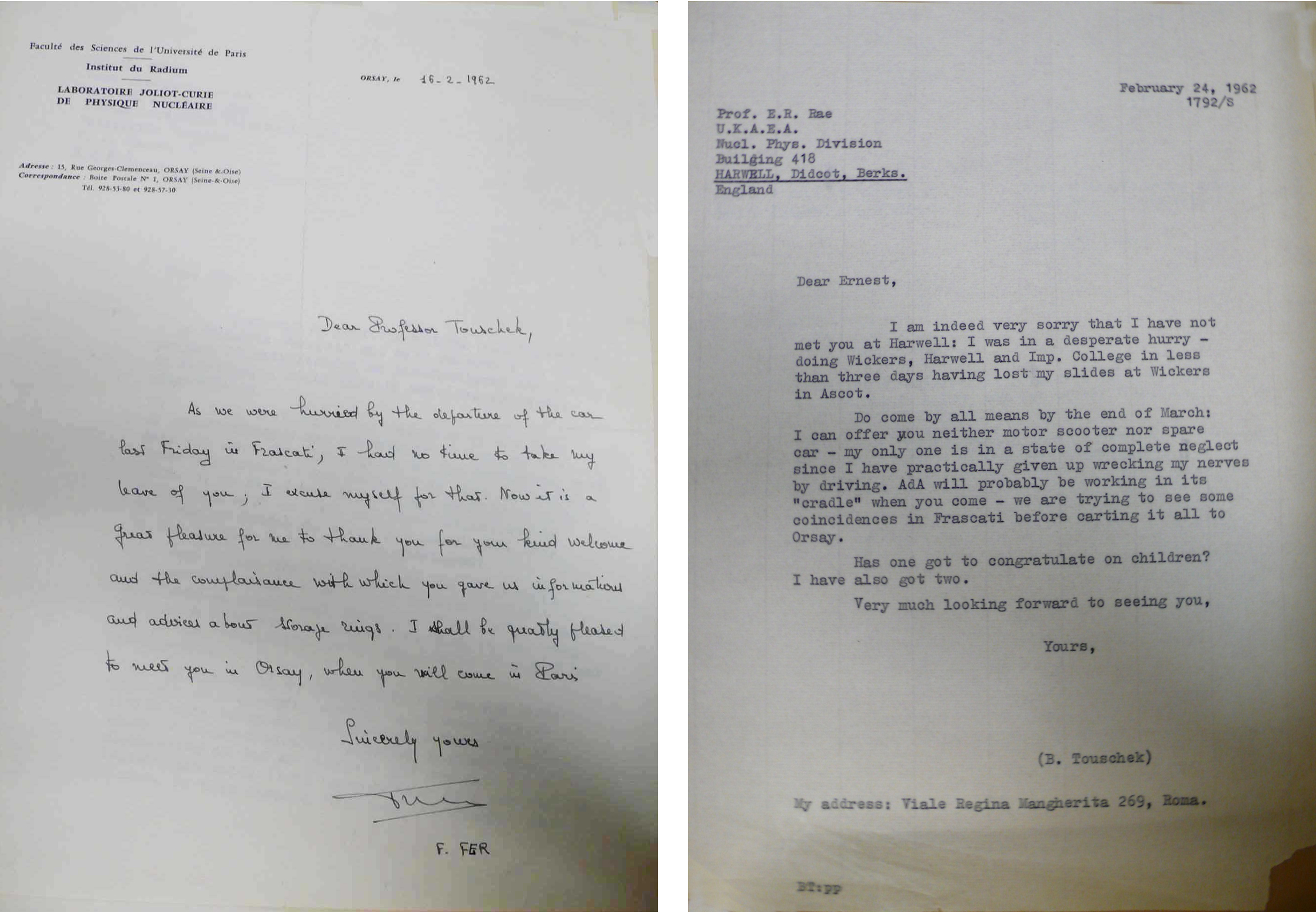}
 \caption{February  16th, 1962 letter from F. Fer to Bruno Touschek thanking him after visiting Frascati at left, and  February 24th, 1962 letter from Touschek to E. R. Rae, at the United Kingdom Atomic Energy Authority, indicating a visit to Harwell by Touschek (Bruno Touschek Archive).}
 \label{fig:FerToTouschek-Rae}
  \end{figure}

  % \begin{figure}[t]
   %[htb]
  %\centering
 % \includegraphics[scale=0.167]{FerATouschek} \hspace{.5cm}
%  \includegraphics[scale=0.15]{BT-RAE-P1120937}
%  \caption{\textcolor{blue}{February  16th, 1962 letter from F. Fer to Bruno Touschek thanking him after visiting Frascati at left, and  February 24th, 1962 letter from Touschek to E. R. Rae, at the United Kingdom Atomic Energy Authority, indicating a visit to Harwell by Touschek.} (Archive Bruno Touschek, Sapienza University of Rome).}
% \label{fig:FerToTouschek-Rae}
%  \end{figure}
   
 By the end of February, Touschek was already confident that AdA would go to Orsay, at the same time he was also actively pursuing approval for the bigger ADONE project. 
 %\textcolor{red}
 {Now that the ``ice had been broken'', as Touschek wrote in April to Blanc-Lapierre (see Fig.~\ref{fig:TouschekALapierre}),} more visits and exchanges between Frascati and Orsay soon followed. The letters which passed  between the two Laboratories during this period  are very warm and friendly, as both sides were now  eager to pursue the possible transfer of AdA  and discuss the practical implications. 
 
  \begin{figure}[htb]
  \centering
  \includegraphics[scale=0.19]{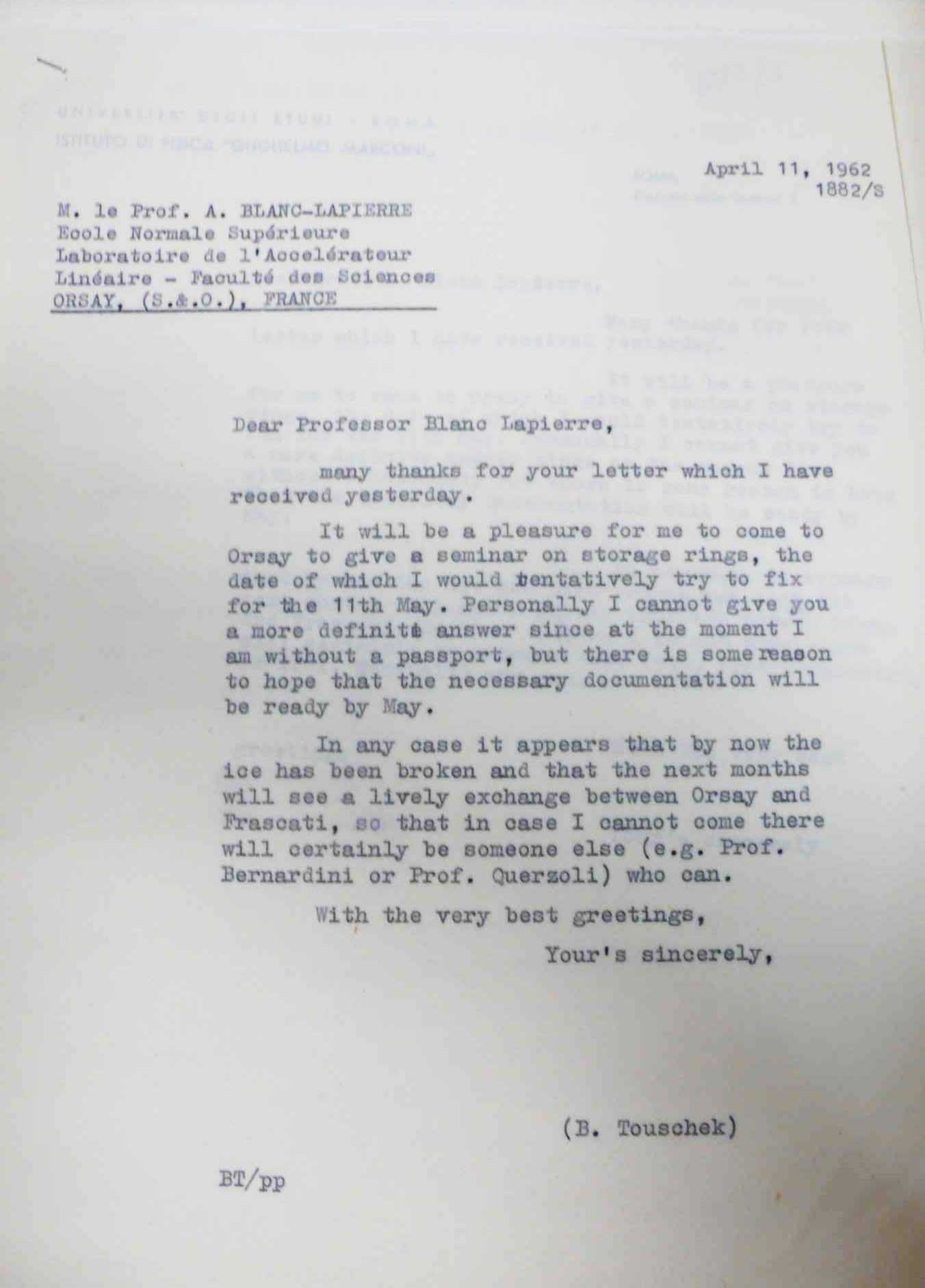} 
\caption{%\textcolor{red}
{Copy of the} letter from Bruno Touschek to Andr\'e Blanc-Lapierre (Bruno Touschek Archive).}
 \label{fig:TouschekALapierre}
  \end{figure}
   
        Interest in the Frascati projects was  not limited to the French Laboratory. In February, Touschek was invited to attend a meeting at CERN organized by Kjell Johnsen,  to examine large accelerator projects  for the future of high energy physics.\footnote{The subject of the meeting appears in the April 4th letter by  \ABL\  mentioned 
  later in the text. For info about Johnsen, see \url{http://cerncourier.com/cws/article/cern/31195}.}
        % \textcolor{red}
        {Since 1961, a special Study Group on New Accelerators led by Johnsen had been established to work out the design study of the Intersecting Storage Rings, a proton-proton collider which would be the second large CERN machine \citep{Russo:1996ab}. Johnsen was clearly quite interested in Touschek's participation to the meeting.}
        The invitation was very pressing, as we can see from Touschek's answer:\footnote{B.T. to K. Johnsen, 21 February 1962, Bruno Touschek Archive, Series 1, Folder. 4, Box 1, 1962-67 Correspondence.}
     \begin{quote}
    Dear Dr. Johnsen,
    
    your threat of ringing me has been transmitted by Amman and I am looking forward to its execution.
    
    I shall certainly come for 3 or 4 days to participate at the enthusiast's meeting \dots
    
    With many greetings and looking forward to your call \dots
     \end{quote}
    %However, at the end Touschek did not participate to the meeting.
    %, which was instead attended by Amman and Bernardini, who met with \ABL \ in this occasion.   
    \ABL \ had been   looking ahead to a possible encounter with Touschek in Geneva  at the end of March, in the context of this    meeting organized by Kjell Johnsen. However,  notwithstanding his earlier positive response,  
    %,  to examine large accelerator projects  for the future of high energy physics \textcolor{blue}{\bf to check}. \footnote{http://cerncourier.com/cws/article/cern/31195} 
    for some reasons, Touschek did not attend the meeting. But  Carlo Bernardini and Amman were there and  the practical details  of AdA's  transfer to Orsay  started to be discussed.  \ABL \ was sorry not to have met Touschek, and pressed him, on an April 4th letter, to come to Orsay and give a seminar.\footnote{Bruno Touschek Archive, Series 1, Folder. 4, Box 1, 1962-67 Correspondence.} \BT \ welcomed the invitation, but had no  passport at the time. Hoping to receive it in time,   he  chose the date of May 11th for his visit.\footnote{Bruno Touschek Archive, Series 1, Folder. 4, Box 1, 1962-67 Correspondence.}

  %\textcolor{red}
  It is uncertain  whether Touschek was able to go to Paris and Orsay for the promised seminar. However, there exists  a testimony of Touschek's  visiting   Paris and giving a seminar,  from  the theoretical physicist Maurice L\'evy, professor at \ENS\, who had been instrumental in calling \ABL\ from Algiers to  France in 1961.
  %L\'evy's  anecdote  shines  a beam of light over Bruno's person outside Italy, but Levy, 
  In telling the episode, L\'evy is uncertain about the precise dating, tentatively placing it in the 1950's.\footnote{This anecdote is reported in \citep{Bernardini:2015wja}, and was originally  told to one of us, G.P., by Maurice L\'evy during a interview in Paris, on May 24th, 2013.}}    
  It is tempting to place the visit  in  May  1962,  and because L\'evy's anecdote    brings  Bruno's personality into  a vivid light,   this is  what happened:
      \begin{quote} I knew Touschek by reputation and I had met him at several conferences. In Paris in %\textcolor{red}
      { the framework of} our theory group, we had a weekly seminar where we invited people from all over and  on several occasions we invited Touschek, who came and talked to us. Unfortunately I do not remember on which subject he talked. In fact he came at least two or three times. \dots I have a small anecdote. In one of the visits, we had put him up in a small hotel on Boulevard Saint Michel  and before retiring at night, he had put his shoes outside the door of his room, and when he opened the door on the next day, the shoes had disappeared. So,  he had a problem, the owner of the hotel kept saying ``this is a small hotel sir, we don't make shoes" and so on. Finally the proprietor of the hotel lent him a pair of shoes, which were much too big for him, two or three sizes too big, and he went to a shop, at nine o'clock when the shops open, to get another pair of shoes for himself. He told us the story later on, [when he came to the Laboratory,] with great sense of humour, Touschek was well known for his sense of humour.
\end{quote}
    The months to follow saw many lively exchanges, and detailed plans for AdA's transfer were made. One of the members of the  team of the Linear Accelerator,  Fran\c cois Lacoste, remembers visiting Frascati in the Spring of 1962:\footnote{
 From an interview with \FL \  on May 2013, in Orsay.} 
   \begin{quote}
   I was witness to some  initial contacts between Frascati and Orsay in  '62 and I had occasion of one visit to Frascati with Boris Milman, where I met Bernardini and Touschek and we looked at some details about  how it would be possible to bring AdA to Orsay. 
   \end{quote}
   \section{How AdA left Italy and arrived in France}
%\textcolor{red}
%\paragraph{HOW ADA LEFT  ITALY AND ARRIVED IN  FRANCE}
{The transfer posed many technical problems, such as  maintaining  the vacuum in the doughnut as AdA would travel across the 1500 kilometres and more between Frascati and Orsay. A major challenge in the  storing of  positrons was in fact  the requirement of an extreme vacuum in the doughnut, to prevent scattering with the residual gas. To this aim, a legendary   vacuum as low as  $5 \times 10^{-10}$mmHg was reached in Frascati by Gianfranco Corazza, one of the physicists in the AdA team, together with Angelo Vitale, a  technician  who specialized in  outgassing in very low pressure vessels. Reaching  such a low level  vacuum required two or three months, and it was  essential not to  lose it during the transfer. As  the vacuum was maintained by powerful devices constantly pumping  residual gas from the doughnut,    batteries would be needed to  keep  the pumps working and  accompany  AdA    through Italy and the   Alps  into France and  on to Orsay.
Two trucks were hired for the transfer: a bigger one with AdA, sealed with its vacuum, with pumps and batteries, and a smaller one for other heavy equipment. Other lighter components could follow by plane. 

%\textcolor{red}
{There were also problems of a different nature: the head of the Italian National  Committee for Nuclear Energy (CNEN), a government agency owning and overseeing the Frascati Laboratories, needed to be contacted and his agreement for AdA to leave Italy had to be obtained. Touschek, advised by Amaldi, wrote a letter and AdA  was granted permission to go. Another possible problem involved the crossing of the border between Italy and France. The chosen route was across the Alps and the custom station was at Modane, near the Frejus.\footnote{%\textcolor{red}
{As the Frejus road tunnel did not yet exists, roads would go through the Moncenisio Pass.}} Passing the French customs could be tricky, and had to be prepared in advance. Both Italian and French officers in the foreign ministries needed  to be informed so as to act in case of difficulties.} 

%\textcolor{red}
When the time  for the transfer came near, Touschek sent  a letter to  Francis Perrin, Haut Commissaire \`a l'\'Energie Atomique  au CEA, who had worked with  \FJ \ on nuclear chain reactions,  and  was very powerful and influential  on all  science matters in France. Perrin had been one of the founders of CERN, and knew Amaldi very well. Informal exchanges between them started as the date of transfer drew close.  At the end of June, everything was finally ready. 

In Fig.~\ref{fig:TouschekToPerrin} we show Touschek's letter of June 28th, 1962, \footnote{Bruno Touschek Archive, Series 1, Folder. 4, Box 1, 1962-67 Correspondence.} which 
 reads:  
    
     \begin{quote}
     Dear Professor Perrin,
     
     I enclose a list of material for the second convoy Frascati-Orsay, which will presumably leave Rome on the 4th of  July and should arrive in Paris on the 7th.
     
     We very much hope that there will be no difficulties at the customs but, in case of emergency, we would much appreciate the help so kindly offered by you to Prof. E. Amaldi.
     
    [\dots] it contains the vacuum chamber at $5\times 10^{-10}$ mm \dots The ideal solution would be if some competent official at the Modane customs office could be informed before hand.
     
     I will take the liberty of wiring you the exact (as near as possible) time at which the convoy can be expected to pass the frontier [\dots]

     \end{quote}
     
     \begin{figure}[htb] 
     \centering \includegraphics[scale=0.52]{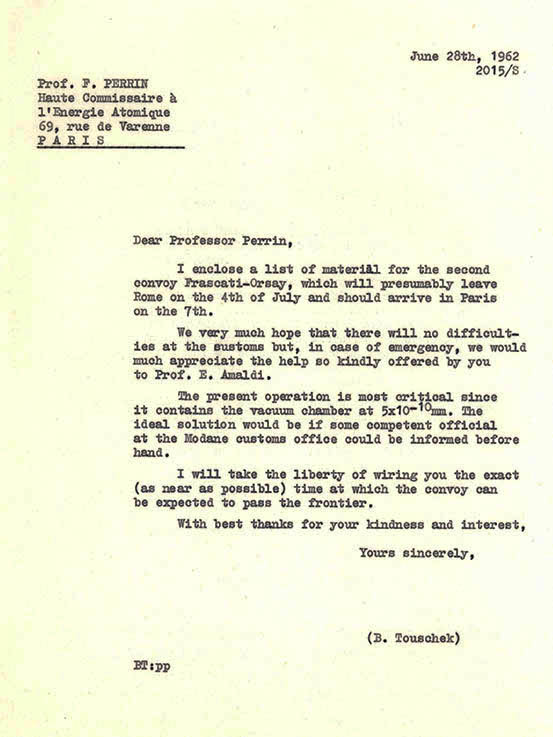} 
     \caption{Copy of the letter from Bruno Touschek to Francis Perrin about the transfer of AdA to Orsay (Bruno Touschek Archive).} \label{fig:TouschekToPerrin}
       \end{figure} 
  
%\paragraph{TO FRANCE: CROSSING THE BORDERS INTO FRANCE }
  The  story of  AdA's transfer  from Italy to France has been told  on various occasions  by its protagonists, Carlo Bernardini, who saw the convoy leave Frascati, and    Fran\c cois Lacoste, who saw it arrive in Orsay.\footnote{Carlo Bernardini, {\it Fisica vissuta}, 2006 Codice Edizioni, Torino,  and Fran\c cois Lacoste  interviewed  by GP, May 2013,  Orsay.} 
  
  \begin{figure}[htb]
  \centering
  \includegraphics[scale=0.4]{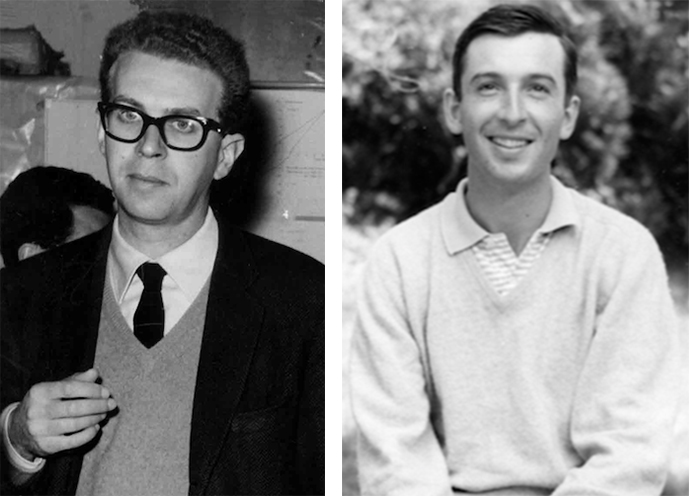}
%  {carloBernardini}\hspace{0.2cm}  \includegraphics[scale=0.5]{lacoste}
\caption{Carlo Bernardini at left,  in the 1960's. At right a photo of \FL, taken in St
Jean de Luz, during  August 1962, the  month  just following  AdA's arrival in Orsay.
{Photos courtesy of Carlo Bernardini and \FL.}}
 \label{fig:carloLacoste}
  \end{figure}
When the time came for the transfer, the Frascati Laboratories called a trusted moving firm, in the person of the  legendary Signor Grossi, as Bernardini calls him, who was often employed by the Laboratories  for moving equipment to and from University of Rome or others.\footnote{{\it Il mitico signor Grossi}, in Bernardini's words.} The 8  ton  iron   magnet inside which AdA's doughnut was placed,  was put on the truck,  with its set of batteries to keep the pumps working, and maintain the vacuum. Any mishap,  namely any air that got into the doughnut would have meant months of extra work in Orsay to clean the chamber. The batteries had to be able to provide power for two, maximum three  days, enough to cover the trip from Frascati to Orsay.
  
There was some worry that such weight could unbalance the truck and make it difficult to keep control. Signor Grossi did not think so, but Touschek was very concerned. So he jumped on the driver's seat and started driving the truck around the large  square in front  of the synchrotron building. Perhaps he just wished to  try  the driving, the result instead was to destroy a lamp post. Properly subdued, he let Signor Grossi take over and finally AdA left. 

Finally  the truck reached the state border between Italy and France, in Modane, on the French side of the Alps. Angelo Vitale, the specialist} of the vacuum, was travelling with the convoy. Customs checks were a very serious business at the time, so the customs officer asked:`` What's inside?", meaning inside the bulky green round object, out of which protruded a short  tube with a small round glass window. This was the window through which scientists and their visitors in  Frascati could see the synchrotron light emitted by the circulating electrons. What could Vitale say? The legend goes that he answered ``There is nothing, just nothing", which was indeed the truth. But this was not sufficient  to let AdA pass. So the help which Touschek had asked for, in his letter to Perrin, became necessary. Vitale called Corazza, who  called Amaldi, who called Perrin, who called the French Minister of Interior, and from there, down the line of command to the customs officer,  until   finally AdA was allowed to enter  France.   
   
The story as told by Lacoste is very similar:
 \begin{quote}
 The transfer of AdA to Orsay came during the summer  of 1962 and I remember waiting for AdA and witnessing the arrival of AdA in Orsay. 
 
 We had to wait a bit longer than we thought because AdA was stopped at the {\it fronti\`ere}  by the French customs, who wanted to understand what they were bringing, what the team from Frascati was bringing and they were especially suspicious of what was inside  AdA, and they wanted to look into, perhaps thinking that it could be drugs or whatever, into the vacuum ring, which was pumped during the trip because they had made the degassing, and so on, and it was vacuum of $10^{-8}$ and at that time good vacuum was very difficult to bring, so they had the pump working during the transit. So the customs  wanted to open it and they only had a small window to ask [to be  opened]. So they asked what the window was, and  the Frascati people answered ``It is in sapphire" , which didn't improve the situation because sapphire is a jewel, for customs.\footnote{Carlo Bernardini, one of the physicists who built AdA,  says  the window was  not made of sapphire, but of quartz. Private communication to the authors.} They had to wait and find a solution and,  luckily, we had the intervention of Francis Perrin. \dots 
 \end{quote}

 AdA and the two trucks arrived in Orsay, and the empty trucks left, to go back to Italy. The image is still vividly recalled in Lacoste's words:
\begin{quote} 
 I remember the trucks going back and, as the small truck was empty and the  big truck also, they decided  that it was easier for       the driver of the truck to put the small truck on the big truck and I remember seeing them travel back that way.  
 \end{quote}
 
 %\textcolor{red}
 
 AdA's ``wedding trousseau", i.e. {\it il corredo}, as Carlo Bernardini used to call it in Italian,  was completed a few days later  with the remaining equipment being sent by plane. The list, detailed in a July 6th  letter by Carlo Bernardini to Fran\c cois Lacoste, reads like a nursery rhyme:\footnote{Copy of this letter is courtesy of Mario Fascetti, the AdA  technician who had built AdA's  radiofrequency system, under Mario Puglisi's supervision.}
 % \textcolor{blue}{\bf Lia: origine di questa  lettera?}
 % \begin{quote}
  \begin{description}
  \item Two Cherenkov glass counters \dots
  \item  one oscillograph \dots
  \item  one Movie camera \dots
  \item  12  power supplies \dots
  \item  30 modular units \dots
  \item 6 scintillation counters
  \item \dots
  \end{description}
  %\end{quote} }

 %\textcolor{red}
 {And then, after AdA and its outfits had arrived, the scientists  and the 
technicians came. AdA was installed in Salle 500 next to the Linear Accelerator, and a new team, now composed of French and Italian scientists, moved  on to  the last leg of AdA's great adventure.\footnote{"Salle 500" means Salle 500 MeV,
500 Mev being the electron beam energy which was delivered by the
Linac in this experimental hall.} In Orsay, the Italian team joined forces with \FL, \PM\   and a young graduate, \JH, who would prepare his {\it Th\`ese d'\'Etat} on AdA, completed three years later, under the guidance of \ABL.
% Pierre Marin, Carlo Bernardini and Bruno Touschek. 
 In Fig.~\ref{fig:JH-PM-BT} we show a photo of \JH\  with  \PM, %\textcolor{red}
 {taken
 % \sout{in Orsay} 
 in the early 1960s}.
%, and, in the right panel,  Bruno Touschek, after a class,  at  University of Rome.}
 \begin{figure}
 %[htb]
  \centering
  \includegraphics[scale=1.6]{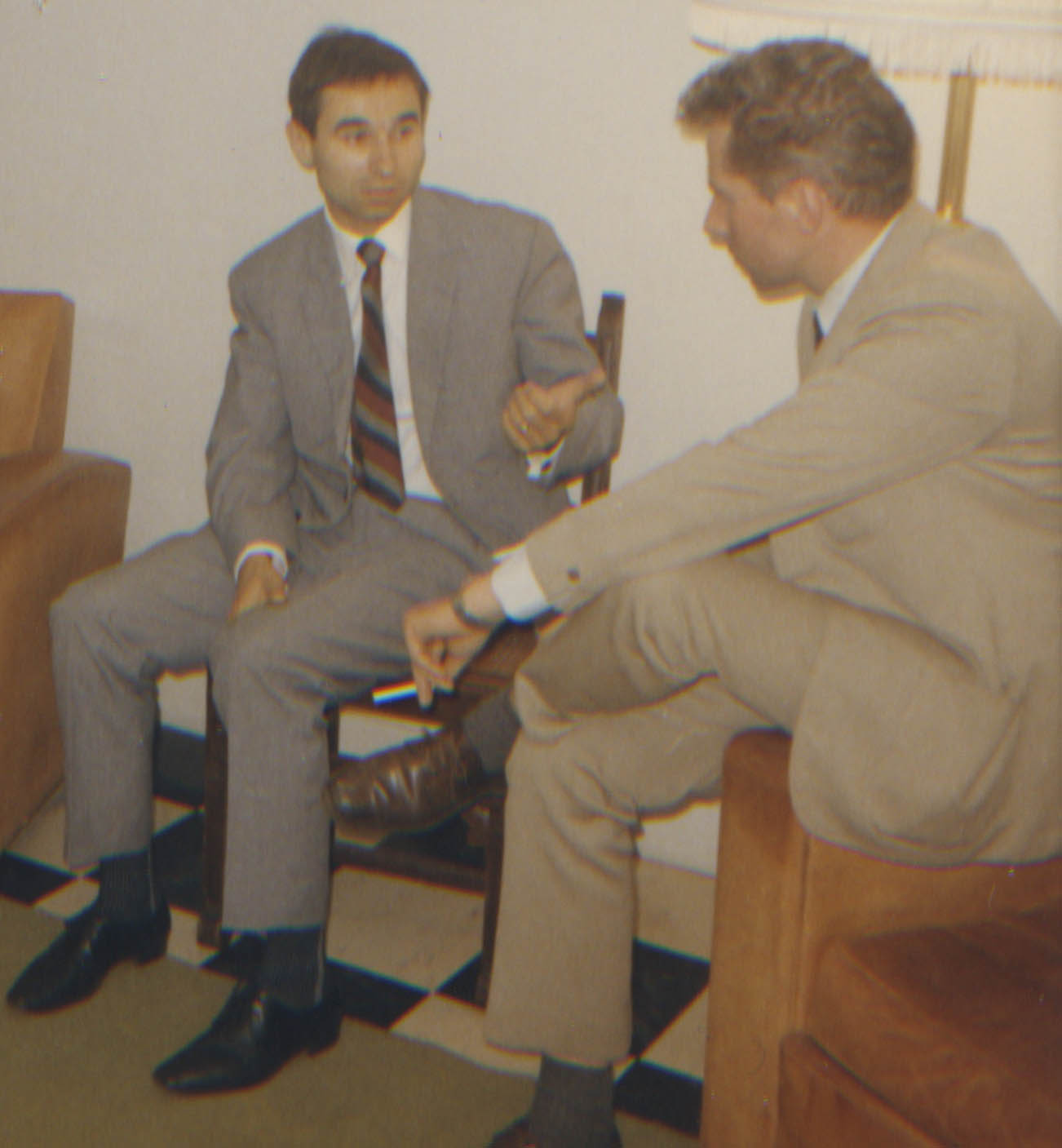}
\caption{\JH \ (at left) and \PM \  in the early 1960s (courtesy of \JH).}
%AdA installed in the Salle de Cible at Orsay Laboratories in  \textcolor{red}{photo courtesy of \JH,} 
 \label{fig:JH-PM-BT}
  \end{figure}

Thus  the adventure of AdA in Orsay started.

\section{What happened in Orsay}
In Orsay, the Italian scientists found the 
ideal conditions to successfully show the feasibility of electron-positron colliders as best suited to probe   the structure  and dynamics of the world of elementary particles.\footnote{Carlo Bernardini used to say that they  soon understood they had been dealt a very god deal, {\it un buon affare}, in Italian.}  Ada allowed  to discover unsuspected collective effects between colliding bunches of particles, and,  measuring  them, to anticipate how future machines should be constructed to exploit  the potential of colliding beam physics for future discoveries. The main observations  from AdA's experimentation concern collective effects about life time and beam size in storage rings, such as  the so-called {\it Touschek effect} \citep{Bernardini:1997sc}. The Touschek effect, discovered in 1963,  is still relevant in planning today's colliders, notwithstanding the enormous magnification in size and energy of present day colliders with respect to  little AdA.

The discovery of  the Touschek effect led to  understand  the correct size of the volume occupied by the particles when the two bunches of oppositely charged particles met, and  allowed the Franco-Italian team  to establish that collisions among electrons and positrons had taken place \citep{Haissinski:1998aa}.    It is appropriate to quote here \JH, from an interview which took place in May 2013, at \LAL, in Orsay:
\begin{quote}
 Les mesures finales qui ont \'et\'e faites ont
 port\'e sur le nombre de collisions par seconde
entre les \'electrons et les positrons. C'\'etait la premi\`ere fois au
monde que l'on montrait que les particules, effectivement, interagissaient et
entraient en collision les unes avec les autres et donc, \c ca montrait que, on peut dire, que ces machines \'etaient utilisables pour faire de la physique
des tr\`es hautes \'energies, et l'essentiel, pas tout, bien s\^ur, on a fait d'autres
d\'ecouvertes par la suite, mais l'essentiel, quand m\^eme, des caract\'eristiques de ce type de machines \'etait valid\'e 
 et permettait de penser que les g\'en\'erations ult\'erieures seraient utilisables pour faire de la physique des particules \`a tr\`es haute \'energie.\footnote{The final measurements relied on  the rate  of collisions among electrons and positrons.   It was the first time in the world that one could show that the particles effectively  interacted and collided against each other, and this allowed to think  that these machines could be used to do [experiments] in  high energy physics. There we other discoveries, of course, but the essential point was that the characteristics of this type  of machines  were validated and that successive generations [of these machines] could  be used for  very high energy physics .  }
\end{quote}  

Thus AdA and its team of Franco-Italian scientists laid the grounds for the  ``fantastic progress" which Richard Feynman 
had unknowingly divined  in his address at the Aix-en-Provence  conference only 
ten months before AdA reached Orsay.\footnote{In 1965 Richard Feynman was awarded the Nobel Prize in Physics  for the  "fundamental work in quantum electrodynamics, with deep-ploughing consequences for the physics of elementary particles", jointly to Sin-Itiro Tomonaga and  Julian Schwinger.}

\section{Conclusions}

In this paper  we have described how French and Italian scientists in the early 1960's created a collaboration which brought the first  electron-positron particle collider AdA, Anello di Accumulazione, storage ring in English, from Frascati to Orsay, a transfer crucial to the success of AdA as a precursor of future colliders. The events which we have described are part of a larger historical scenario, in which different European countries,  following separate ways before, through, and after World War II, %\textcolor{red}
{ prepared  the scientific and technological conditions for AdA to be built in Frascati} and collisions between electrons and positrons to be observed in Orsay \citep{Bernardini:1964lqa}.

 The present  paper   is part of a work in progress for  a project,  in which the various European roads to particle colliders will be  described in detail. This project will   include an account of the circumstances which brought Bruno Touschek and the Norwegian  \RW\ to work together  in Germany, during WWII, on a secret project for the construction of a betatron, financed by the Ministry of Aviation of the Third Reich. It is during this period that \RW \ envisaged the possibility to construct an accelerator able to  make collisions between particles of opposite signs, an idea he shared with \BT, and later patented \citep{%Waloschek:1994qp}.
 Wideroe:1994}. 
 As Touschek himself always acknowledged, this was the first time he had ever heard of such idea. When the times became ripe for transforming Wider\o e's idea into a working machine, twenty years later,  Touschek remembered the long ago conversations he had with \RW\  in Berlin,  as they were planning the betatron and the  war was entering its darkest period, and together with his collaborators proposed and started to build AdA. Of course, to transform the idea into reality, many things had to happen, among them     the construction of the Orsay Linear Accelerator  and the Frascati electron synchrotron, both of which will  be described in the project under preparation.

  We are now posting this part of the AdA story   in memory of Bruno Touschek, the Austrian born theoretical physicist who was the  prime mover behind AdA and who passed away, 40 years ago, on May 25th, 1978, in Innsbruck, Austria. 
  %\textcolor{red}
  {We close  this preliminary presentation of our work in progress}  with  Fig.~\ref{fig:BT}}, a  photograph of  Bruno Touschek taken in the University of Rome, around 1960. 
   \begin{figure}[htb]
  \centering
    \includegraphics[scale=1.5]{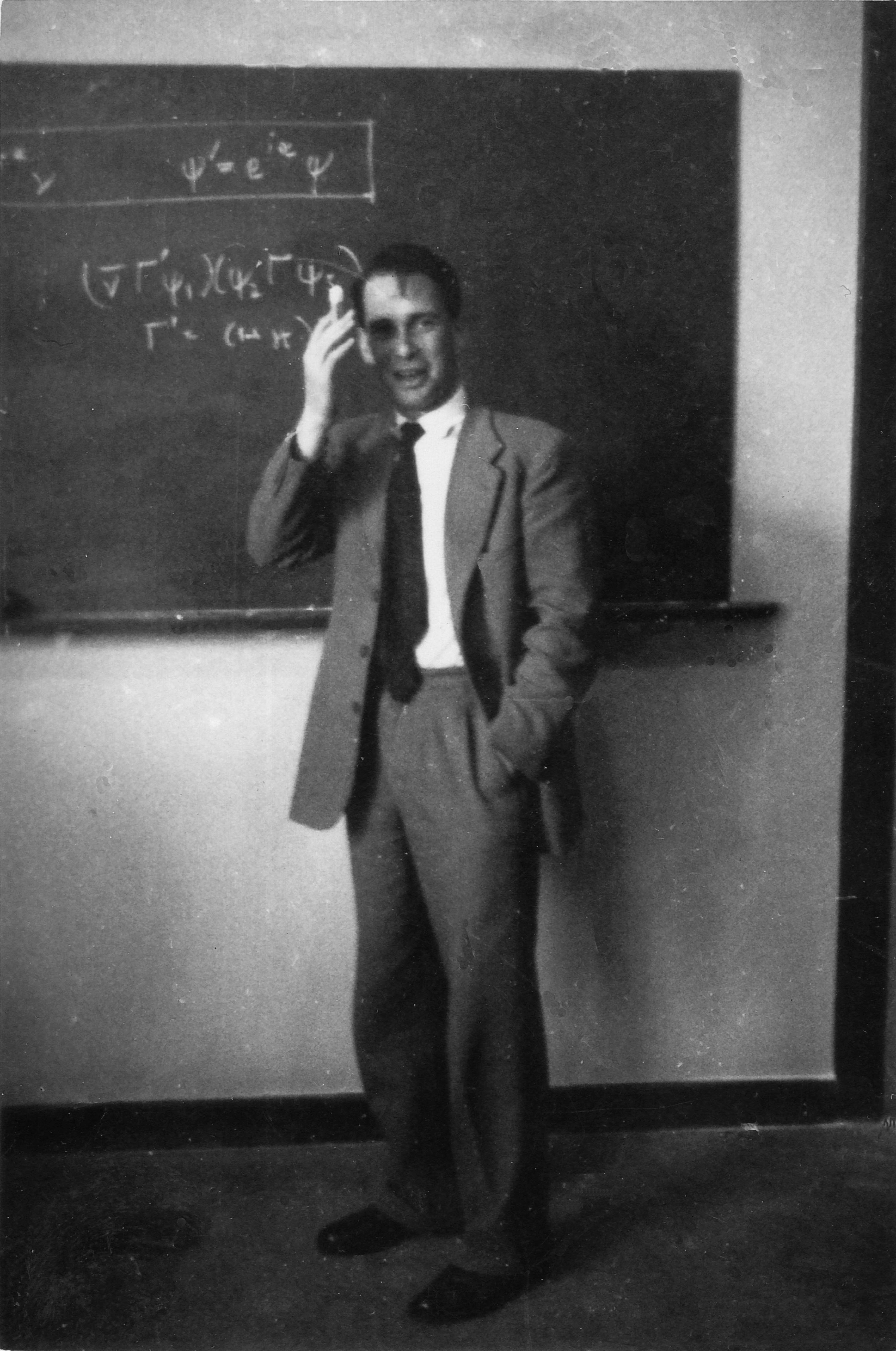}
\caption{
%\textcolor{red}
{Bruno Touschek at University of Rome, around 1960 (courtesy of Francis Touschek).}}
 \label{fig:BT}
  \end{figure}
%\par\vskip 1 cm
\section*{Acknowledgements}
We thank Jacques Ha\"issinski for invaluable advice on the early AdA history and a critical reading of the manuscript. We gratefully acknowledge his many useful suggestions and  help  in providing  archival documents on the exchanges between Rome and Paris, which were to lead to AdA's transfer to France. 
One of us, G.P., is grateful to Earle Lomon for  encouragement and criticisms, and gratefully acknowledges hospitality at the MIT Center for Theoretical Physics, while this article was completed. We thank the \LAL\ for hospitality during the May 2013 interviews quoted here. We  thank Orlando Ciaffoni, Antonio Cupellini and Claudio Federici,  from INFN Frascati National Laboratories, 
%Corrado Mencuccini, Alessandro Pascolini and Vincenzo Valente 
for providing support, library assistance and access to photographs. %\textcolor{red}
We also thank Giovanni Battimelli  
%\textcolor{red}
and the Sapienza University of Rome for access % \textcolor{red}
{to the Edoardo Amaldi Archives of the Department of Physics and for allowing reproduction of documents from Bruno Touschek Papers}.
{We  thank Francis Touschek for kind permission to use \BT's photos.}

\bibliographystyle{aipauth4-1}
%unsrt}
%plain}
%epjh-year}
\bibliography{Touschek_june6}
\end{document}